\documentclass[ALICE,manyauthors,onecolumn]{cernphprep}
\usepackage[comma,square,numbers,sort&compress]{natbib}
\usepackage{hyperref}
\usepackage{lineno}
\usepackage{multirow}
\usepackage{amsmath}
\usepackage{siunitx}
\DeclareSIUnit\eVperc{\eV\per\clight}
\DeclareSIUnit\clight{\text{\ensuremath{c}}}

\newcommand{\pim}{\ensuremath{\pi^{-}}} 
\newcommand{\pt}{\ensuremath{p_{\rm T}}\,} 
\newcommand{\pT}{\ensuremath{p_{\rm T}}~} 
\newcommand{\tXi}{\ensuremath{\Xi}\,}

\newcommand{\La}{\ensuremath{\Lambda}\,}

\newcommand{\aLa}{\ensuremath{\overline{\Lambda}}\,}
\newcommand{\pp}{\ensuremath{\mathrm {p\kern-0.05em p}}\,}
\newcommand{\ppNoSpace}{\ensuremath{\mathrm {p\kern-0.05em p}}}
\newcommand{\pPb}{\ensuremath{\mbox{p--Pb}}\,}
\newcommand{\xim}{\ensuremath{\Xi^{-}}\,}
\newcommand{\xip}{\ensuremath{\overline{\Xi}^{+}}\,}

\newcommand{\pP}{\ensuremath{\mbox{p--p}}~}

\newcommand{\pXim}{\ensuremath{\mbox{p--$\Xi^{-}$}}~}
\newcommand{\pXip}{\ensuremath{\overline{\mathrm{p}}$\mbox{--}$\overline{\Xi}^+}~}

\newcommand{\NLa}{\ensuremath{\mbox{N--$\Lambda$}}~}

\newcommand{\NXi}{\ensuremath{\mbox{N--$\Xi$}}~}
\newcommand{\NSig}{\ensuremath{\mbox{N--$\Sigma$}}~}

\newcommand{\pXi}{\ensuremath{\mbox{p--$\Xi$}}~}

\newcommand{\minScoul}{\ensuremath{3.6}\,}
\newcommand{\maxScoul}{\ensuremath{5.3}\,}
\newcommand{\minShal}{\ensuremath{1.8}\,}
\newcommand{\maxShal}{\ensuremath{3.7}\,}
\newcommand{\radiusResult}{\ensuremath{r_0 = 1.427 \pm 0.007 \mathrm{(stat.)}^{+0.001} _{-0.014}\mathrm{(syst.)}}~}

\newcommand{\MeVc}{\ensuremath{\mathrm{MeV}\kern-0.05em/\kern-0.02em \textit{c}}~}
\newcommand{\GeVc}{\ensuremath{\mathrm{GeV}\kern-0.05em/\kern-0.02em \textit{c}}~}
\newcommand{\GeVcSq}{\ensuremath{\mathrm{GeV}\kern-0.05em/\kern-0.02em \textit{c}^2}~}
\newcommand{\MeVcSq}{\ensuremath{\mathrm{MeV}\kern-0.05em/\kern-0.02em \textit{c}^2}~}

\newcolumntype{b}{X}
\newcolumntype{s}{>{\hsize=.5\hsize}X}

\begin{document}%

\begin{titlepage}
\PHyear{2019}
\PHnumber{080}      
\PHdate{25 April}  
%
\title{First observation of an attractive interaction between a proton and a cascade baryon}
\ShortTitle{First observation of an attractive interaction between proton and $\Xi$ baryons}

\Collaboration{ALICE Collaboration\thanks{See Appendix~\ref{app:collab} for the list of Collaboration members}}
\ShortAuthor{ALICE Collaboration} 

\begin{abstract}
This work presents the first experimental observation of the attractive strong interaction between a proton and a multi-strange baryon (hyperon) $\Xi^-$. The result is extracted from two-particle correlations of combined \pXim$\oplus$ \pXip pairs measured in \pPb collisions at $\sqrt{s_{\mathrm{NN}}}=5.02$\,TeV at the LHC with ALICE.
The measured correlation function is compared with the prediction obtained assuming only an attractive Coulomb interaction and a standard deviation in the range  $[\minScoul, \maxScoul] $ is found. Since the measured  \pXim$\oplus$ \pXip correlation is significantly enhanced with respect to the Coulomb prediction, the presence of an additional, strong, attractive interaction is evident.
The data are compatible with recent lattice calculations by the HAL-QCD Collaboration, with a standard deviation in the range $ [\minShal, \maxShal]$. The lattice potential predicts a shallow repulsive $\Xi^-$ interaction within pure neutron matter and this implies stiffer equations of state for neutron-rich matter including hyperons. Implications of the strong interaction for the modeling of neutron stars are discussed.

\end{abstract}
\end{titlepage}
\setcounter{page}{2}

Hyperons are baryons containing at least one strange quark (e.g. $\Lambda=uds$, $\Sigma^0=uds$, $\Xi^-=ssd$) and hyperon-nucleon interactions are the object of intensive studies for two main purposes. The first one is to achieve a level of precision in the strangeness sector of low-energy Quantum Chromodynamics (QCD) comparable to the one reached in the determination of the scattering parameters of nucleon-nucleon interactions. The second purpose is to study the impact of the strong interaction betweeen baryons with strangeness on the description of dense objects within astrophysics \cite{Lattimer:2004pg,Djapo:2008au,SchaffnerBielich:2000wj,Haidenbauer:2016vfq}.\\
Effective field theory provides a systematic expansion scheme to compute hyperon-nucleon and hyperon-hyperon interactions \cite{Haidenbauer:2013oca,Haidenbauer:2016vfq} but currently the experimental constraints are rather scarce.\\
Scattering experiments~\cite{SechiZorn:1969hk,Eisele:1971mk,Alexander:1969cx} and spectroscopy of several hypernuclei~\cite{Hashimoto:2006aw} established the attractive character of the \NLa interaction but only scarce information is available for \NSig~\cite{PhysRevLett.80.1605,Hayano:1988pn} and \NXi~\cite{Nagae:2017slp,Nakazawa:2015joa} interactions. \\
Hyperon-nucleon (p$\Lambda$, p$\Omega$) and hyperon-hyperon ($\Lambda\Lambda$) interactions were already investigated by means of two-particle correlations in the momentum space measured in heavy-ion collisions by the STAR collaboration \cite{Adams:2005ws,Adamczyk:2014vca,STAR:2018uho}. However, these analyses are hampered by large statistical uncertainties or by contamination by non-genuine contributions to the correlation function \cite{Morita:2014kza}, and hence new experimental approaches are called for.

Recently it has been shown that hyperon-nucleon, hyperon-hyperon \cite{Ourpaper,Acharya:2019yvb} and kaon-nucleon \cite{Acharya:2019bsa} interactions can be more precisely measured in proton-proton (\ppNoSpace) and proton-lead (p-Pb) collisions at the LHC. Indeed, small colliding systems at LHC energies lead to particle-emitting sources with sizes of about \SI{1}{fm}, allowing a precise test of the short-range strong interaction. With an emitting source size similar to that of \pp collisions~\cite{Adam:2015pya}, the larger number of pairs available in the data set recorded from \pPb collisions at $\sqrt{s_{\mathrm{NN}}}= 5.02\,$TeV by ALICE allows these studies to be extended to the \pXim correlation. The newly developed tool CATS (Correlation Analysis Tool using the Schr\"odinger equation)~\cite{Mihaylov:2018rva} allows to compute predictions for the the \pXim correlation considering either only the known Coulomb interaction or including additionally a strong potential. The direct comparison of the measured and predicted correlation functions provides an unprecedented tool to test the strong \pXi interaction.

In this work we present the first evidence of a strong attractive interaction in the \pXim channel. We also compare the experimental correlation to the prediction obtained employing lattice calculations from the HAL-QCD Collaboration \cite{HAL1,HAL2} for the \pXim interaction. This, but also any  other \pXim potential, can be then used to evaluate the single-particle potential of the $\Xi^-$ within pure neutron matter \cite{Inoue:2016qxt}.
The possible appearance of $\Xi^-$ within dense neutron matter depends on this  single-particle potential ~\cite{Weissenborn:2011kb}. An attractive single-particle potential for the $\Xi^-$ within pure neutron matter would favor the appearance of $\Xi^-$  at already moderate densities~\cite{Annala:2017llu}, softening the Equation of State (EoS), while a repulsive single-particle potential~\cite{SchaffnerBielich:2000wj} would shift the $\Xi^-$ production to larger densities~\cite{Haidenbauer:2016vfq} and stiffen the EoS.\\
These studies are relevant for the modelling of neutron stars since, due to the large densities achieved in the center of these objects, neutrons might transform into hyperons to minimize the system energy \cite{Lonardoni:2014bwa}. 
So far primarily $\Lambda$ hyperons are included in  theoretical calculations because the $\Lambda$-nucleon interaction is better known than the $\Xi$-nucleon and $\Sigma$-nucleon interactions, but all the three hyperon-species and their interactions with nucleons should be considered to achieve a realistic equation of state.

It is clear that the precise measurement of the \pXim strong interaction will allow for a sound determination of the corresponding single-particle potential and consequently for more realistic EoSs of neutron stars with hyperon content.

This letter presents $\mbox{p--p}\oplus\overline{\mathrm{p}}\mbox{--}\overline{\mathrm{p}}$ and $\mbox{p--$\Xi^-$}\oplus\overline{\mathrm{p}}\mbox{--}\overline{\Xi}^+$ correlations measured in \pPb collisions at $\sqrt{s_{\mathrm{NN}}}=5.02$\,TeV employing the data set collected by ALICE~\cite{1748-0221-3-08-S08002, Abelev:2014ffa} in 2016 during the LHC Run 2. As the correlation functions of baryon-baryon pairs exhibit identical behavior compared to their respective anti-baryon\mbox{--}anti-baryon pairs~\cite{Adam:2015vja,Adamczyk:2015hza}, the corresponding samples are combined. Therefore, in the following \pP denotes the combination of $\mbox{p--p} \oplus \overline{\mathrm{p}}\mbox{--}\overline{\mathrm{p}}$, and accordingly for \mbox{p--$\Xi^{-}$}. Collision events are triggered by the coincidence in the V0 scintillator arrays~\cite{1748-0221-8-10-P10016}, which is also used to reject background events stemming from interactions of the beam particles with the beam-pipe materials or beam gas. 
Pile-up events with more than one \pPb collision per bunch crossing are rejected by evaluating the presence of multiple event vertices. To assure a uniform detector coverage, the distance along the beam axis between the reconstructed primary vertex and the nominal interaction point is required to be smaller than $10$\,cm. After these selection criteria are applied, about $600 \times10^{6}$ minimum-bias events are available for the analysis.

The main detectors used in the analysis are the Inner Tracking System (ITS)~\cite{1748-0221-3-08-S08002} and the Time Projection  Chamber (TPC)~\cite{2010TPCNIMA}, covering the full azimuthal angle and the pseudorapidity range of $|\eta| < 0.9$. These detectors are located within a solenoid that creates a magnetic field of $B =$ \SI{0.5}{T} directed along the beam axis. The measurement of the specific ionization energy loss, d$E$/d$x$, in the TPC gas, and the time information delivered by the Time of Flight (TOF)~\cite{Akindinov2013} detector are used for particle identification (PID). Particles originating from weak decays are differentiated from primary\cite{ALICE-PUBLIC-2017-005} particles originating at the collision point 
since their associated tracks do not point to the primary vertex
\cite{Abelev:2014ffa}.  

The 
proton candidates are identified following the same criteria listed in~\cite{Ourpaper}. The TPC and TOF PID capabilities are used to select 
protons by the deviation of the PID signal from its expectation value normalized to units of standard deviations $n_{\sigma,\mathrm{proton}}$ of the detector resolution ($\sigma_{\mathrm{TPC}}$, $\sigma_{\mathrm{TOF}}$). DPMJET~\cite{DPMJET} Monte Carlo events processed such as to emulate the ALICE detector acceptance and reconstruction algorithm~\cite{1748-0221-3-08-S08002} are used to estimate the purity and composition of the selected samples. Both proton and antiproton samples are found to have a purity of 97\%, and to consist of 86\% primary particles.

The $\tXi^-$ baryons are reconstructed~\cite{Adam:2015vsf} using the decay channel $\xim \rightarrow \La\pim$\cite{Tanabashi:2018oca}. The \La is identified by its decay channel $\La\rightarrow \mathrm{p}\pim$~\cite{Tanabashi:2018oca}. The charged particles employed in the $\tXi^-$ reconstruction are selected via PID with $\mid n_{\sigma,\mathrm{TPC,i}}\mid<4$ ($i=\pi$,p), and they are required to have a hit in one of the ITS layers or a matched TOF signal in order to use timing information to remove the contribution of particles stemming from out-of-bunch pile-up. The \La candidates are selected by applying the following topological criteria: i) a minimum distance for the \La daughter tracks to the primary vertex of \SI{0.05}{cm}, ii) a maximum distance between the two daughter tracks of \SI{1.5}{cm} iii) the radial distance of the \La decay vertex to the detector center in radial coordinates, $r_{\mathrm{xy}}$, in the range \num{1.4} to \SI{200}{cm}, and iv)  the cosine of the pointing angle (CPA) between the \La momentum and the vector connecting the primary and decay vertices is required to be $\mathrm{CPA} > 0.97$.

The \La invariant mass is calculated using the pion and proton hypothesis for the daughters and is described by a double Gaussian, accounting for the signal and the mass resolution, and a second-order polynomial for the combinatorial background. The resulting average mass resolution is \SI{2.0}{MeV\per\square\clight} independent of transverse momentum (\pt) of the selected candidates. A total of \num{18.0e6} (\num{17.6e6}) \La (\aLa) candidates are selected within $\pm3\sigma$ around the nominal mass, with a signal ($S$) to background ($B$) ratio $S/B$ of 5.1 (5.4) corresponding to a purity of 83.5\% (84.3\%).

A $\pi^-$ candidate track is combined with the selected \La candidate to form a $\xim$ and evaluate its decay vertex. The following topological selection criteria are applied: i) a minimum distance for the $\pi^-$ to the primary vertex of \SI{0.05}{cm}, ii) a maximum distance between the track of the $\pi^-$ and the $\Lambda$  of \SI{1.5}{cm}, iii) a $r_{\mathrm{xy}}$ of the \xim decay vertex between \num{0.8} and \SI{ 200}{cm}, and iv) a minimum \xim CPA of \num{0.98}. 
The \xim mass resolution increases from \SI{2.1}{MeV\per\square\clight} at low \pT to \SI{2.7}{MeV\per\square\clight} at larger \pt, with a \pT averaged value of \SI{2.3}{MeV\per\square\clight}. Applying a $\pm2\sigma$ selection of the average value around the nominal \xim mass, a $S/B$ ratio of 7.3 (7.9), resulting in purities of \SI{87.9}{\percent} (\SI{88.6}{\percent}), is estimated for \xim (\xip). A total of \num{8e5} $\Xi$ candidates of each charge are selected.
%
The fraction of primary particles is calculated considering measured production rates of $\Omega$~\cite{ALICE:2017jyt} and  $\Xi^{0}(1530)$~\cite{Xi1530}, and assuming for the $\xim(1530)$ a similar production rate as for the $\Xi^{0}(1530)$.
The total sample is hence estimated to consist of 66.1\% primary particles. 

%
Experimentally, the correlation function is computed as
$C(k^*)=\mathcal{N}\frac{A(k^*)}{B(k^*)},$
where $k^*=\frac{1}{2}\cdot|\mathbf{p}^*_1-\mathbf{p}^*_2|$ is the reduced relative momentum of two particles with momenta $\mathbf{p}^*_1 \text{ and }\mathbf{p}^*_2$ in the pair rest frame ($\mathbf{p}^*_1 =-\mathbf{p}^*_2$), $A(k^*)$ represents the same event $k^*$ distribution, and $B(k^*)$ is a corresponding reference sample of uncorrelated pairs obtained by pairing particles from different events \cite{Ourpaper}. The normalization constant $\mathcal{N}$ between the two distributions is obtained in the region $k^*\in$~[\num{240},\num{340}]~\si{MeV\per\clight}, where final state interaction effects are absent and the correlation function is flat. The theoretical correlation function $C(k^*) = \int S(\textbf{r}) |\psi_{k^*}(\textbf{r})|^2 \textrm{d}^3r$ in this work is computed with CATS~\cite{Mihaylov:2018rva}, where \textbf{r} is the relative distance between the two particles, $S(\textbf{r})$ is the source function and $\psi_{k^*}(\textbf{r})$ is the two-particle wave function. A spherically-symmetric emitting source with a Gaussian density profile parameterized by a radius parameter $r_0$ is assumed and Coulomb and strong potentials are considered to evaluate the relative wave functions for \pP and \pXim pairs.

\begin{figure*}[t]
  \centering
  \includegraphics[width=0.49\linewidth]{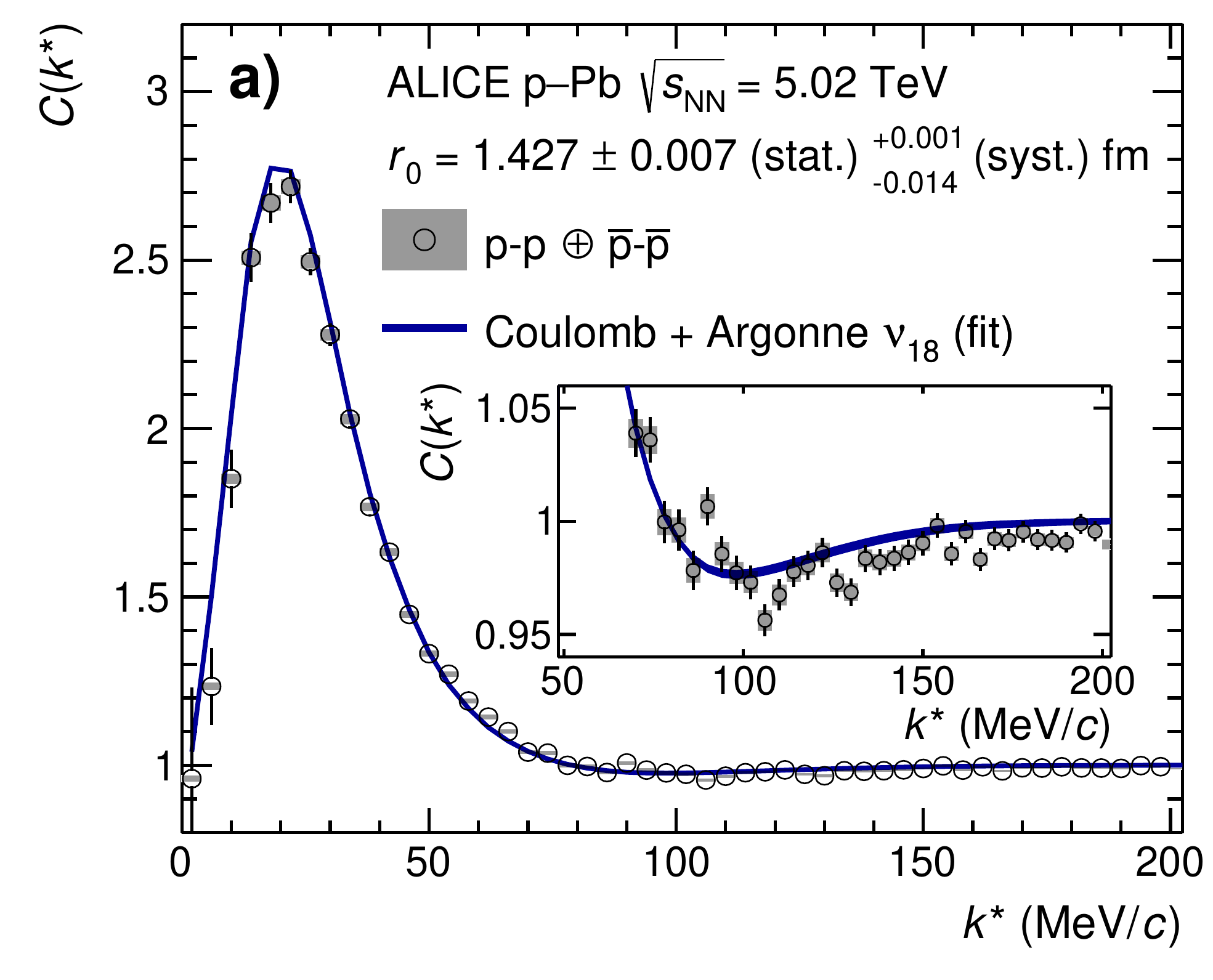}
  \hfill
  \includegraphics[width=0.49\linewidth]{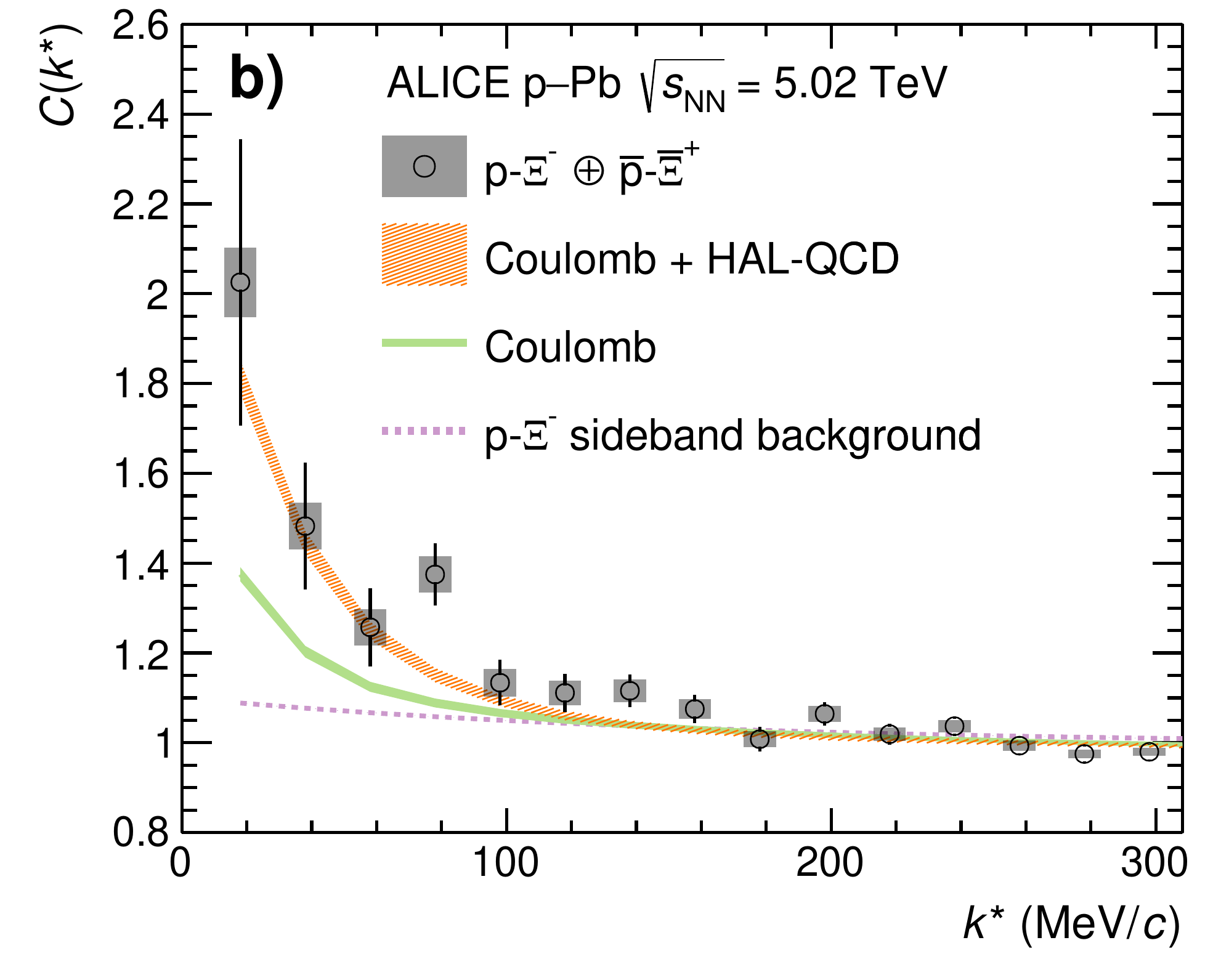}
\caption{\textit{(Color online)} The a) \pP and b) \pXim correlation functions shown as a function of $k^*$. Statistical (bars) and systematic uncertainties (boxes) are shown separately. The filled bands denote the results from the fit with Eq.~\ref{eq:lambdaCF}. Their widths correspond to one standard deviation of the systematic error of the fit. The HAL-QCD curve uses potentials obtained from
\cite{Hatsuda:2018nes}. The dashed line in the right panel shows the contribution from misidentified \ensuremath{\mbox{p--$\tilde{\Xi}^{-}$}}\,pairs from the sidebands scaled by its $\lambda$ parameter. See text for details.}
 \label{fig:CF}
\end{figure*}

The measured correlation functions for \pP 
and \pXim
are shown in Fig.~\ref{fig:CF}.
The inset in the left panel shows a zoom of the \pP correlation function around $k^* =$ \SI{100}{MeV\per\clight}, where the effect of the repulsive interaction can be seen. 
A total number of \num{574e3} (\num{412e3}) $\mbox{p--p}$ ($\overline{\mathrm{p}}\mbox{--}\overline{\mathrm{p}}$) and \num{3.3e3} (\num{2.6e3}) $\mbox{p--$\Xi^-$}$ ($\overline{\mathrm{p}}\mbox{--}\overline{\Xi}^+$) pairs contribute to $A(k^*)$ in the region $k^*$\SI{<200}{MeV\per\clight}. The systematic uncertainties for the \pP and \pXim correlations are obtained by varying all single-particle selection criteria for protons and $\Xi$ candidates with respect to their default values such as to obtain a maximum variation of the single particle yields of $\pm15$\% . The resulting uncertainties on the correlation functions are symmetrized and added in quadrature.\\  In order not to be dominated by statistical fluctuations, the systematic uncertainties are evaluated in intervals of \SI{40}{MeV\per\clight} width in $k^*$ for \pP and \SI{200}{MeV\per\clight} for p--$\Xi^{-}$, and fitted by a second order polynomial which serves to interpolate the final point-by-point correlated uncertainties in narrower intervals.
The total systematic uncertainty reaches a maximum value of \SI{5}{\percent} for \pP and \SI{3.2}{\percent} for p--$\Xi^{-}$ at the lowest measured $k^*$ value.

The experimental data are fitted with the model correlation function obtained from CATS, $C_{\mathrm{model}}(k^*)$. Together with the genuine correlation function due to the two-particle interaction, residual correlations are also considered. In the experiment the latter are introduced by contamination of the selected samples due to particle misidentification and feed-down from weak decays of other particles. These are taken into account according to
\begin{equation}
C_{\mathrm{model}}(k^*) = 1 + \lambda_{\mathrm{genuine}} \cdot ( C_{\mathrm{genuine}}(k^*)-1)+\sum_{\mathrm{ij}}\lambda_{\mathrm{ij}}(C_{\mathrm{ij}}(k^*) -1),
\label{eq:lambdaCF}
\end{equation}
where $C_{\mathrm{genuine}}(k^*)$ is the genuine correlation function for the pairs of interest, $i$ and $j$ denote all possible impurity and feed-down contributions, and $C_{\mathrm{ij}}(k^*)$ represent the corresponding correlation functions. The parameters $\lambda_{\mathrm{ij}}$ are the relative weights of these contributions calculated from  purity and feed-down fractions \cite{Ourpaper} and are summarized in Tab.~\ref{tab:lambdaval}. Here $\tilde{X}$ denotes misidentified particles and $X_Y$ particles originating from the decay of $Y$. Both the \pP and \pXim correlation functions are dominated by the genuine correlation of interest. The main contribution contaminating the \pP correlation function are protons from \La or $\Sigma^+$ weak decays. The genuine \pXim signal is diluted with contributions from secondary protons as mentioned above, misidentified $\Xi\mathrm{s}$, or from decays of the $\Xi(1530)$ resonance. For the feed-down contributions, the shape of the $C_{\mathrm{ij}}(k^*)$ correlations is obtained by transforming the initial theoretical correlation function \cite{HAIDENBAUER201324} of the mother particles via the corresponding decay matrices \cite{PhysRevC.89.054916}. For most combinations this results in a flat $C_{\mathrm{ij}}(k^*)\sim 1$. For contributions with misidentified particles a flat correlation is assumed except for the case of p--$\tilde{\Xi}^-$, where experimental data from the side-bands of the invariant mass selection are used. This contribution is also shown in Fig.~\ref{fig:CF} after scaling according to $1+\lambda_{\rm{p{-}}\tilde{\Xi}^-} (C_{\rm{p{-}}\tilde{\Xi}^-}(k^*)-1)$.
\begin{table*} 
\begin{center}
\caption[lambda parameters]{Weight of the individual components of the \pP and \pXim correlation function. Entries in the form $X_Y$ denote particles originating from the decay of $Y$, whereas $\tilde{X}$ denotes misidentified particles. Non-flat contributions are listed individually.}
\begin{tabularx}{.9\textwidth}{ l|X || l|X  }
\hline  \hline
\multicolumn{2}{c||}{\pP} & 
\multicolumn{2}{c}{\pXim} \\

Pair & $\lambda$ parameter [\%] &
Pair & $\lambda$ parameter [\%] \\

\hline 
\pP  & 72.1 &
\pXim & 51.3  \\

p--p$_{\Lambda}$ & 16.1 &
p--$\Xi^-_{\Xi^- (1530)}$ & 8.2   \\

Feed-down (flat) & 8.7&
p--$\tilde{\Xi}^-$ & 8.5  \\

Misidentification (flat) & 3.1 &
Feed-down (flat) & 29.1 \\

 &  &
Misidentification (flat) & 2.9 \\

\hline \hline
\end{tabularx}
\label{tab:lambdaval}
\end{center}
\end{table*}

\begin{figure}[ht]
  \centering
  \includegraphics[width=0.95\linewidth]{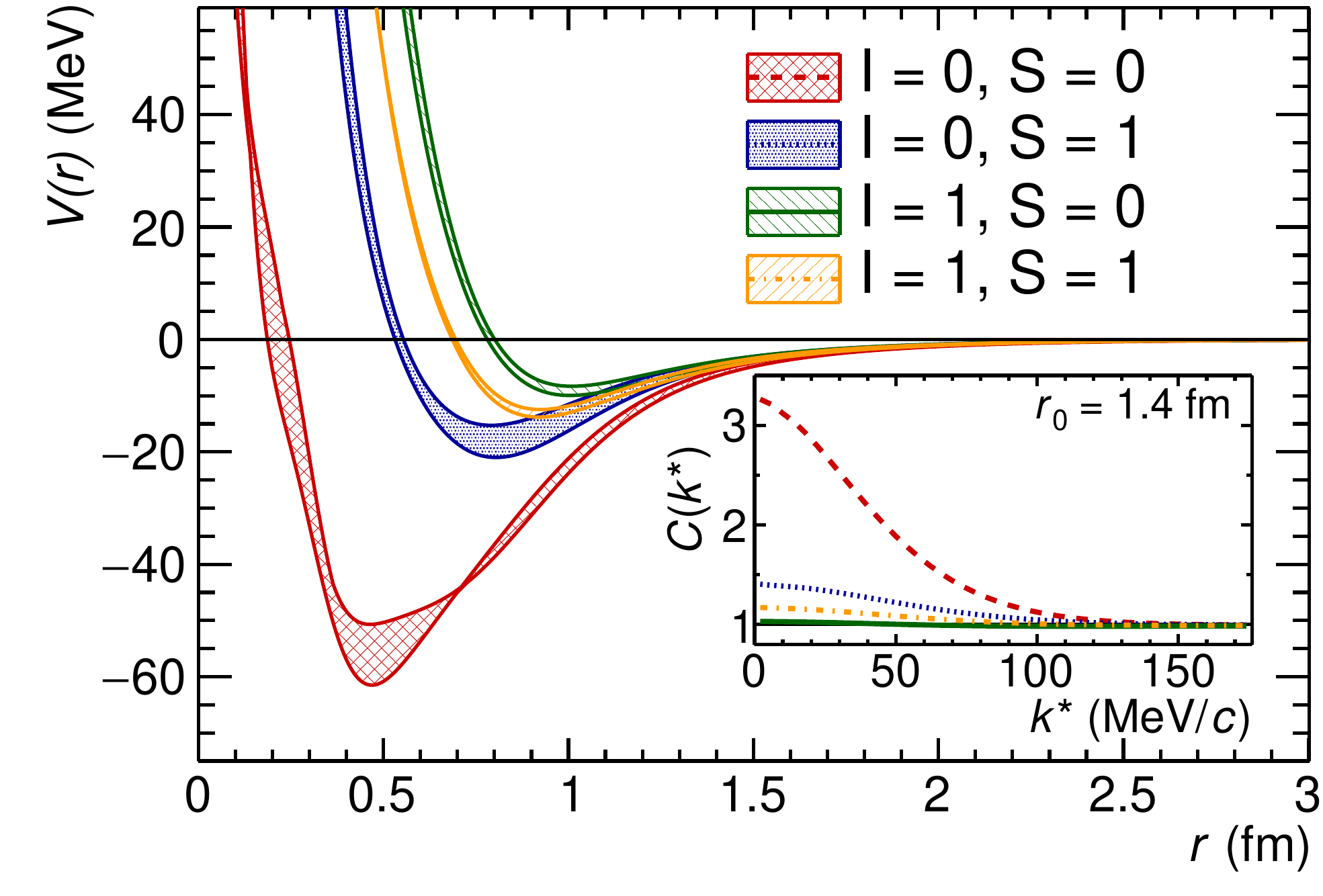}
  \caption{\textit{(Color online)}~Predictions for the $\Xi$-nucleon potential from the HAL-QCD Collaboration
  \cite{Hatsuda:2018nes}
  for the different spin (S) and isospin (I) states.  The error bands refer to different Euclidean times considered in the calculation. The inset shows the correlation function computed with the central value of the potential for each of the different states and a source radius of \SI{1.4}{fm}.
  }
  \label{fig:pXiPot}
\end{figure}
The genuine \pP correlation function 
is computed
by using the Coulomb and the strong  Argonne $v_{18}$~\cite{PhysRevC.51.38} potentials, considering $s$ and $p$ waves. The radius $r_0$ of the emitting source is a free parameter determined by a fit to the experimental \pP correlation function, conducted in $k^*\in$~[\num{0},\num{375}]~\si{MeV\per\clight}. A normalization parameter $a$ is included for the final fit function to the data $C_{\mathrm{\mathrm{tot}}}(k^*)$ in the form $C_{\mathrm{\mathrm{tot}}}(k^*) = a \cdot C_{\mathrm{model}}(k^*)$, and it is also determined by the fit, driven by the flat region extending from \SI{200}{MeV\per\clight}. The theoretical correlation is smeared to account for the finite momentum resolution.  \\
Although Fig.\,\ref{fig:CF} shows that no mini-jet background is visible for baryon-baryon correlations~\cite{Adam2017,Ourpaper},
possible deformations of the correlation function due to energy and momentum conservation were considered by extending the fit procedure. A systematic variation of the fit
is carried out by adding a baseline $C_{\rm non\text{-}femto}(k^*)$
in the form
$C_{\mathrm{\mathrm{tot}}}(k^*) = C_{\rm \mathrm{non\text{-}femto}}(k^*) \cdot C_{\mathrm{\mathrm{model}}}(k^*) = (a+b\cdot k^*)\cdot C_{\mathrm{model}}(k^*)$.
The parameters $a$ and $b$
are estimated from the fit to the \pP data.
Additional systematic uncertainties of the fit and of the radius $r_0$ are evaluated by varying: i) the range of the fit region up to $350$ or \SI{400}{MeV\per\clight}, and
ii) the $\lambda$ parameters by modifying the secondary contributions by $\pm20$\% while keeping the sum of the primary and secondary fractions constant. The widths of the filled bands in Fig.~\ref{fig:CF} correspond to one standard deviation of the total systematic error of the fit. 

The resulting radius \radiusResult \si{fm} obtained by a fit with a $\chi^2/\mathrm{ndf} = 1.42$ is then used in the computation of the \pXim correlation function, following the premise of a common Gaussian source. Differences in the multiplicity dependence of the radius for \pP and \pXim pairs have been investigated and found to be negligible. For the \pXim interaction, two scenarios were tested: one considering only the Coulomb interaction and a second one with an additional strong potential computed on the lattice and provided by the HAL-QCD Collaboration
\cite{Hatsuda:2018nes}.

Figure \ref{fig:pXiPot} shows the $\Xi$-nucleon strong interaction potential as a function of the pair separation distance $r$ for the different combinations of isospin ($\mathrm{I}= 0,\,1$) and spin ($\mathrm{S}= 0, \,1$). The widths of the potentials correspond to the uncertainties of the lattice calculations. The inset shows the correlation functions computed with the average values of each component of the potential and for a source radius equal to \SI{1.4}{fm}. The different correlation functions obtained for the four I, S channels show the sensitivity  to \pXim distances lower than \SI{1.5}{fm}. Nevertheless, a precise test of the potential for small distances will be possible only by improving the statistical uncertainties of the measurement by a factor of 10, as expected during the LHC Run 3. \\
The genuine total \pXim correlation is obtained by computing the correlation function
including the Coulomb and strong interaction for the four different states with CATS and then summing up the correlation functions with their specific
statistical weights,
\begin{equation}
\begin{split}
C_{\pXim} = &~\frac{1}{8} C_{\NXi}(\mathrm{I}=0,\,\mathrm{S}=0) + \frac{3}{8} C_{\NXi}(\mathrm{I}=0,\,\mathrm{S}=1) \\
& + \frac{1}{8} C_{\NXi}(\mathrm{I}=1,\,\mathrm{S}=0) +\frac{3}{8} C_{\NXi}(\mathrm{I}=1,\,\mathrm{S}=1).
\end{split}
\end{equation}

The computation of the \pXim correlations is carried out by first fitting the normalization parameter $a$ in the range $k^*\in$~[\num{250},\num{600}]~\si{MeV\per\clight}, where the correlation function is flat. Then, using the  resulting $C_{\rm tot}(k^*)$, the correlation function is compared with experimental data.

Systematic uncertainties of the predicted \pXim correlation function from Coulomb and Coulomb + strong interaction are evaluated by varying: i) the range where the normalization parameter $a$ is estimated to $k^*\in$~[\num{300},\num{550}] and $k^*\in$~[\num{350},\num{700}]~\si{MeV\per\clight}, ii) the fit procedure by including the baseline $C_{\rm non\text{-}femto}(k^*) = (a+b \cdot k^*)$, iii) the $\lambda$ parameters by modifying the secondary contributions by $\pm20$\% while keeping primary and secondary fractions constant, and
iv) the radius $r_0$ by decreasing it by 20\% to account for possible variation of the \pXim source with respect to the \pP source due to the larger contribution of strong $\Delta$ decays to the latter. The theoretical correlation is smeared to account for the finite momentum resolution and its width in Fig.~\ref{fig:CF} corresponds to one standard deviation of the total systematic uncertainty in the model evaluation.

The comparison of the experimental \pXim data with the predicted correlation functions including only the Coulomb potential and the Coulomb + strong potential  in Fig.~\ref{fig:CF} shows that the latter is favored. The fact that the experimental \pXim correlation function shows a stronger enhancement than the Coulomb-only assumption is able to produce means that the total interaction is more attractive than the  assumption of a Coulomb-only interaction. The exclusion of this scenario is quantified by computing the p-value of the data-model comparison considering statistical and systematic errors. To account for the systematic errors of the experimental data, the yield in each $k^*$ bin is smeared according to a Gaussian distribution with a width equal to the systematic error of each bin and all obtained permutations are compared to the Coulomb- only and Coulomb + strong correlation functions. 
 The obtained p-values are converted into $n_{\sigma}$ values.
The Coulomb-only correlation function is compared with the data in $k^*\in$~[\num{0},\num{140}]~\si{MeV\per\clight} and the obtained $n_{\sigma}$ distributions present a standard deviation   from \minScoul to \maxScoul. For the Coulomb + strong interaction, the $n_{\sigma}$ values range from \minShal to \maxShal. The observation of a significant deviation between measured correlation function and the prediction using only the Coulomb interaction provides strong evidence for an attractive strong potential in the \pXi system. 

In order to evaluate the consequences of this new observation for the EoS of neutron stars, the $\Xi^{-}$ single-particle potential in pure neutron matter (PNM) at saturation density from HAL-QCD can be considered. This results in a slight repulsion for $\Xi^{-}$ in PNM of around \SI{6}{MeV}~\cite{Inoue:2016qxt}. Since current 
models~\cite{WEISSENBORN201262} include a much wider range $\in$~[\num{-40},\num{40}]~\si{MeV\per\clight} for such $\Xi^{-}$ single partice potential, the validated lattice predictions impose a much more stringent constraint with consequences for the EoS containing hyperons.
The slight repulsion that the $\Xi^{-}$ single-particle potential acquires in PNM translates into larger densities for the appearance of $\Xi^-$ within neutron-rich matter and into a stiffer EoS. The data to be collected at the LHC in the future will provide the opportunity to study also baryon-antibaryon combinations such as antiproton-$\Xi^-$ correlations.


In summary, this letter presents the first measurement of the \pXim correlation function in \pPb collisions at $\sqrt{s_{\mathrm{NN}}}=5.02$\,TeV. A fit of the \pP correlation function with a model including a quantitative treatment of residual correlations yields a radius of \radiusResult \si{fm} for the emitting source of the particles. The \pXim correlation is compared with Coulomb and Coulomb + strong interaction assumptions and a deviation between \minScoul and \maxScoul $n_\sigma$ to the Coulomb-only correlation is measured. This means that an attractive \pXim strong interaction is observed. The lattice potential provided by the HAL-QCD Collaboration for the \pXim interaction is found to be consistent with our measurements with $n_{\sigma}$ values from \minShal to \maxShal. This measurement constrains models of neutron stars containing hyperons to stiffer EoS. Additional data will allow different models~\cite{Haidenbauer:2018jvl} to be more precisely tested in order to conclude on the presence of $\Xi^-$ hyperons within neutron stars.

\newenvironment{acknowledgement}{\relax}{\relax}
\begin{acknowledgement}
\section*{Acknowledgements}

The ALICE Collaboration would like to thank all its engineers and technicians for their invaluable contributions to the construction of the experiment and the CERN accelerator teams for the outstanding performance of the LHC complex.
The ALICE Collaboration gratefully acknowledges the resources and support provided by all Grid centres and the Worldwide LHC Computing Grid (WLCG) collaboration.
The ALICE Collaboration acknowledges the following funding agencies for their support in building and running the ALICE detector:
A. I. Alikhanyan National Science Laboratory (Yerevan Physics Institute) Foundation (ANSL), State Committee of Science and World Federation of Scientists (WFS), Armenia;
Austrian Academy of Sciences, Austrian Science Fund (FWF): [M 2467-N36] and Nationalstiftung f\"{u}r Forschung, Technologie und Entwicklung, Austria;
Ministry of Communications and High Technologies, National Nuclear Research Center, Azerbaijan;
Conselho Nacional de Desenvolvimento Cient\'{\i}fico e Tecnol\'{o}gico (CNPq), Universidade Federal do Rio Grande do Sul (UFRGS), Financiadora de Estudos e Projetos (Finep) and Funda\c{c}\~{a}o de Amparo \`{a} Pesquisa do Estado de S\~{a}o Paulo (FAPESP), Brazil;
Ministry of Science \& Technology of China (MSTC), National Natural Science Foundation of China (NSFC) and Ministry of Education of China (MOEC) , China;
Croatian Science Foundation and Ministry of Science and Education, Croatia;
Centro de Aplicaciones Tecnol\'{o}gicas y Desarrollo Nuclear (CEADEN), Cubaenerg\'{\i}a, Cuba;
Ministry of Education, Youth and Sports of the Czech Republic, Czech Republic;
The Danish Council for Independent Research | Natural Sciences, the Carlsberg Foundation and Danish National Research Foundation (DNRF), Denmark;
Helsinki Institute of Physics (HIP), Finland;
Commissariat \`{a} l'Energie Atomique (CEA), Institut National de Physique Nucl\'{e}aire et de Physique des Particules (IN2P3) and Centre National de la Recherche Scientifique (CNRS) and Rl\'{e}gion des  Pays de la Loire, France;
Bundesministerium f\"{u}r Bildung, Wissenschaft, Forschung und Technologie (BMBF) and GSI Helmholtzzentrum f\"{u}r Schwerionenforschung GmbH, Germany;
General Secretariat for Research and Technology, Ministry of Education, Research and Religions, Greece;
National Research, Development and Innovation Office, Hungary;
Department of Atomic Energy Government of India (DAE), Department of Science and Technology, Government of India (DST), University Grants Commission, Government of India (UGC) and Council of Scientific and Industrial Research (CSIR), India;
Indonesian Institute of Science, Indonesia;
Centro Fermi - Museo Storico della Fisica e Centro Studi e Ricerche Enrico Fermi and Istituto Nazionale di Fisica Nucleare (INFN), Italy;
Institute for Innovative Science and Technology , Nagasaki Institute of Applied Science (IIST), Japan Society for the Promotion of Science (JSPS) KAKENHI and Japanese Ministry of Education, Culture, Sports, Science and Technology (MEXT), Japan;
Consejo Nacional de Ciencia (CONACYT) y Tecnolog\'{i}a, through Fondo de Cooperaci\'{o}n Internacional en Ciencia y Tecnolog\'{i}a (FONCICYT) and Direcci\'{o}n General de Asuntos del Personal Academico (DGAPA), Mexico;
Nederlandse Organisatie voor Wetenschappelijk Onderzoek (NWO), Netherlands;
The Research Council of Norway, Norway;
Commission on Science and Technology for Sustainable Development in the South (COMSATS), Pakistan;
Pontificia Universidad Cat\'{o}lica del Per\'{u}, Peru;
Ministry of Science and Higher Education and National Science Centre, Poland;
Korea Institute of Science and Technology Information and National Research Foundation of Korea (NRF), Republic of Korea;
Ministry of Education and Scientific Research, Institute of Atomic Physics and Ministry of Research and Innovation and Institute of Atomic Physics, Romania;
Joint Institute for Nuclear Research (JINR), Ministry of Education and Science of the Russian Federation, National Research Centre Kurchatov Institute, Russian Science Foundation and Russian Foundation for Basic Research, Russia;
Ministry of Education, Science, Research and Sport of the Slovak Republic, Slovakia;
National Research Foundation of South Africa, South Africa;
Swedish Research Council (VR) and Knut \& Alice Wallenberg Foundation (KAW), Sweden;
European Organization for Nuclear Research, Switzerland;
National Science and Technology Development Agency (NSDTA), Suranaree University of Technology (SUT) and Office of the Higher Education Commission under NRU project of Thailand, Thailand;
Turkish Atomic Energy Agency (TAEK), Turkey;
National Academy of  Sciences of Ukraine, Ukraine;
Science and Technology Facilities Council (STFC), United Kingdom;
National Science Foundation of the United States of America (NSF) and United States Department of Energy, Office of Nuclear Physics (DOE NP), United States of America.    
\end{acknowledgement}

\bibliographystyle{utphys}   
\bibliography{XiP_femto.bib}


\section{The ALICE Collaboration}
\label{app:collab}

\begingroup
\small
\begin{flushleft}
S.~Acharya\Irefn{org141}\And 
D.~Adamov\'{a}\Irefn{org93}\And 
S.P.~Adhya\Irefn{org141}\And 
A.~Adler\Irefn{org74}\And 
J.~Adolfsson\Irefn{org80}\And 
M.M.~Aggarwal\Irefn{org98}\And 
G.~Aglieri Rinella\Irefn{org34}\And 
M.~Agnello\Irefn{org31}\And 
N.~Agrawal\Irefn{org10}\And 
Z.~Ahammed\Irefn{org141}
% ...
\And
\renewcommand\labelenumi{\textsuperscript{\theenumi}~}

\section*{Affiliation notes}
\renewcommand\theenumi{\roman{enumi}}
\begin{Authlist}
\item \Adef{org*}Deceased
\item \Adef{orgI}Dipartimento DET del Politecnico di Torino, Turin, Italy
\item \Adef{orgII}M.V. Lomonosov Moscow State University, D.V. Skobeltsyn Institute of Nuclear, Physics, Moscow, Russia
\item \Adef{orgIII}Department of Applied Physics, Aligarh Muslim University, Aligarh, India
\item \Adef{orgIV}Institute of Theoretical Physics, University of Wroclaw, Poland
\end{Authlist}

\section*{Collaboration Institutes}
\renewcommand\theenumi{\arabic{enumi}~}
\begin{Authlist}
\item \Idef{org1}A.I. Alikhanyan National Science Laboratory (Yerevan Physics Institute) Foundation, Yerevan, Armenia
\item \Idef{org2}Bogolyubov Institute for Theoretical Physics, National Academy of Sciences of Ukraine, Kiev, Ukraine
\item \Idef{org3}Bose Institute, Department of Physics  and Centre for Astroparticle Physics and Space Science (CAPSS), Kolkata, India
\item \Idef{org4}Budker Institute for Nuclear Physics, Novosibirsk, Russia
\item \Idef{org5}California Polytechnic State University, San Luis Obispo, California, United States
\item \Idef{org6}Central China Normal University, Wuhan, China
\item \Idef{org7}Centre de Calcul de l'IN2P3, Villeurbanne, Lyon, France
\item \Idef{org8}Centro de Aplicaciones Tecnol\'{o}gicas y Desarrollo Nuclear (CEADEN), Havana, Cuba
\item \Idef{org9}Centro de Investigaci\'{o}n y de Estudios Avanzados (CINVESTAV), Mexico City and M\'{e}rida, Mexico
\item \Idef{org10}Centro Fermi - Museo Storico della Fisica e Centro Studi e Ricerche ``Enrico Fermi', Rome, Italy
\item \Idef{org11}Chicago State University, Chicago, Illinois, United States
\item \Idef{org12}China Institute of Atomic Energy, Beijing, China
\item \Idef{org13}Chonbuk National University, Jeonju, Republic of Korea
\item \Idef{org14}Comenius University Bratislava, Faculty of Mathematics, Physics and Informatics, Bratislava, Slovakia
\item \Idef{org15}COMSATS University Islamabad, Islamabad, Pakistan
\item \Idef{org16}Creighton University, Omaha, Nebraska, United States
\item \Idef{org17}Department of Physics, Aligarh Muslim University, Aligarh, India
\item \Idef{org18}Department of Physics, Pusan National University, Pusan, Republic of Korea
\item \Idef{org19}Department of Physics, Sejong University, Seoul, Republic of Korea
\item \Idef{org20}Department of Physics, University of California, Berkeley, California, United States
\item \Idef{org21}Department of Physics, University of Oslo, Oslo, Norway
\item \Idef{org22}Department of Physics and Technology, University of Bergen, Bergen, Norway
\item \Idef{org23}Dipartimento di Fisica dell'Universit\`{a} 'La Sapienza' and Sezione INFN, Rome, Italy
\item \Idef{org24}Dipartimento di Fisica dell'Universit\`{a} and Sezione INFN, Cagliari, Italy
\item \Idef{org25}Dipartimento di Fisica dell'Universit\`{a} and Sezione INFN, Trieste, Italy
\item \Idef{org26}Dipartimento di Fisica dell'Universit\`{a} and Sezione INFN, Turin, Italy
\item \Idef{org27}Dipartimento di Fisica e Astronomia dell'Universit\`{a} and Sezione INFN, Bologna, Italy
\item \Idef{org28}Dipartimento di Fisica e Astronomia dell'Universit\`{a} and Sezione INFN, Catania, Italy
\item \Idef{org29}Dipartimento di Fisica e Astronomia dell'Universit\`{a} and Sezione INFN, Padova, Italy
\item \Idef{org30}Dipartimento di Fisica `E.R.~Caianiello' dell'Universit\`{a} and Gruppo Collegato INFN, Salerno, Italy
\item \Idef{org31}Dipartimento DISAT del Politecnico and Sezione INFN, Turin, Italy
\item \Idef{org32}Dipartimento di Scienze e Innovazione Tecnologica dell'Universit\`{a} del Piemonte Orientale and INFN Sezione di Torino, Alessandria, Italy
\item \Idef{org33}Dipartimento Interateneo di Fisica `M.~Merlin' and Sezione INFN, Bari, Italy
\item \Idef{org34}European Organization for Nuclear Research (CERN), Geneva, Switzerland
\item \Idef{org35}Faculty of Electrical Engineering, Mechanical Engineering and Naval Architecture, University of Split, Split, Croatia
\item \Idef{org36}Faculty of Engineering and Science, Western Norway University of Applied Sciences, Bergen, Norway
\item \Idef{org37}Faculty of Nuclear Sciences and Physical Engineering, Czech Technical University in Prague, Prague, Czech Republic
\item \Idef{org38}Faculty of Science, P.J.~\v{S}af\'{a}rik University, Ko\v{s}ice, Slovakia
\item \Idef{org39}Frankfurt Institute for Advanced Studies, Johann Wolfgang Goethe-Universit\"{a}t Frankfurt, Frankfurt, Germany
\item \Idef{org40}Gangneung-Wonju National University, Gangneung, Republic of Korea
\item \Idef{org41}Gauhati University, Department of Physics, Guwahati, India
\item \Idef{org42}Helmholtz-Institut f\"{u}r Strahlen- und Kernphysik, Rheinische Friedrich-Wilhelms-Universit\"{a}t Bonn, Bonn, Germany
\item \Idef{org43}Helsinki Institute of Physics (HIP), Helsinki, Finland
\item \Idef{org44}High Energy Physics Group,  Universidad Aut\'{o}noma de Puebla, Puebla, Mexico
\item \Idef{org45}Hiroshima University, Hiroshima, Japan
\item \Idef{org46}Hochschule Worms, Zentrum  f\"{u}r Technologietransfer und Telekommunikation (ZTT), Worms, Germany
\item \Idef{org47}Horia Hulubei National Institute of Physics and Nuclear Engineering, Bucharest, Romania
\item \Idef{org48}Indian Institute of Technology Bombay (IIT), Mumbai, India
\item \Idef{org49}Indian Institute of Technology Indore, Indore, India
\item \Idef{org50}Indonesian Institute of Sciences, Jakarta, Indonesia
\item \Idef{org51}INFN, Laboratori Nazionali di Frascati, Frascati, Italy
\item \Idef{org52}INFN, Sezione di Bari, Bari, Italy
\item \Idef{org53}INFN, Sezione di Bologna, Bologna, Italy
\item \Idef{org54}INFN, Sezione di Cagliari, Cagliari, Italy
\item \Idef{org55}INFN, Sezione di Catania, Catania, Italy
\item \Idef{org56}INFN, Sezione di Padova, Padova, Italy
\item \Idef{org57}INFN, Sezione di Roma, Rome, Italy
\item \Idef{org58}INFN, Sezione di Torino, Turin, Italy
\item \Idef{org59}INFN, Sezione di Trieste, Trieste, Italy
\item \Idef{org60}Inha University, Incheon, Republic of Korea
\item \Idef{org61}Institut de Physique Nucl\'{e}aire d'Orsay (IPNO), Institut National de Physique Nucl\'{e}aire et de Physique des Particules (IN2P3/CNRS), Universit\'{e} de Paris-Sud, Universit\'{e} Paris-Saclay, Orsay, France
\item \Idef{org62}Institute for Nuclear Research, Academy of Sciences, Moscow, Russia
\item \Idef{org63}Institute for Subatomic Physics, Utrecht University/Nikhef, Utrecht, Netherlands
\item \Idef{org64}Institute for Theoretical and Experimental Physics, Moscow, Russia
\item \Idef{org65}Institute of Experimental Physics, Slovak Academy of Sciences, Ko\v{s}ice, Slovakia
\item \Idef{org66}Institute of Physics, Homi Bhabha National Institute, Bhubaneswar, India
\item \Idef{org67}Institute of Physics of the Czech Academy of Sciences, Prague, Czech Republic
\item \Idef{org68}Institute of Space Science (ISS), Bucharest, Romania
\item \Idef{org69}Institut f\"{u}r Kernphysik, Johann Wolfgang Goethe-Universit\"{a}t Frankfurt, Frankfurt, Germany
\item \Idef{org70}Instituto de Ciencias Nucleares, Universidad Nacional Aut\'{o}noma de M\'{e}xico, Mexico City, Mexico
\item \Idef{org71}Instituto de F\'{i}sica, Universidade Federal do Rio Grande do Sul (UFRGS), Porto Alegre, Brazil
\item \Idef{org72}Instituto de F\'{\i}sica, Universidad Nacional Aut\'{o}noma de M\'{e}xico, Mexico City, Mexico
\item \Idef{org73}iThemba LABS, National Research Foundation, Somerset West, South Africa
\item \Idef{org74}Johann-Wolfgang-Goethe Universit\"{a}t Frankfurt Institut f\"{u}r Informatik, Fachbereich Informatik und Mathematik, Frankfurt, Germany
\item \Idef{org75}Joint Institute for Nuclear Research (JINR), Dubna, Russia
\item \Idef{org76}Korea Institute of Science and Technology Information, Daejeon, Republic of Korea
\item \Idef{org77}KTO Karatay University, Konya, Turkey
\item \Idef{org78}Laboratoire de Physique Subatomique et de Cosmologie, Universit\'{e} Grenoble-Alpes, CNRS-IN2P3, Grenoble, France
\item \Idef{org79}Lawrence Berkeley National Laboratory, Berkeley, California, United States
\item \Idef{org80}Lund University Department of Physics, Division of Particle Physics, Lund, Sweden
\item \Idef{org81}Nagasaki Institute of Applied Science, Nagasaki, Japan
\item \Idef{org82}Nara Women{'}s University (NWU), Nara, Japan
\item \Idef{org83}National and Kapodistrian University of Athens, School of Science, Department of Physics , Athens, Greece
\item \Idef{org84}National Centre for Nuclear Research, Warsaw, Poland
\item \Idef{org85}National Institute of Science Education and Research, Homi Bhabha National Institute, Jatni, India
\item \Idef{org86}National Nuclear Research Center, Baku, Azerbaijan
\item \Idef{org87}National Research Centre Kurchatov Institute, Moscow, Russia
\item \Idef{org88}Niels Bohr Institute, University of Copenhagen, Copenhagen, Denmark
\item \Idef{org89}Nikhef, National institute for subatomic physics, Amsterdam, Netherlands
\item \Idef{org90}NRC Kurchatov Institute IHEP, Protvino, Russia
\item \Idef{org91}NRNU Moscow Engineering Physics Institute, Moscow, Russia
\item \Idef{org92}Nuclear Physics Group, STFC Daresbury Laboratory, Daresbury, United Kingdom
\item \Idef{org93}Nuclear Physics Institute of the Czech Academy of Sciences, \v{R}e\v{z} u Prahy, Czech Republic
\item \Idef{org94}Oak Ridge National Laboratory, Oak Ridge, Tennessee, United States
\item \Idef{org95}Ohio State University, Columbus, Ohio, United States
\item \Idef{org96}Petersburg Nuclear Physics Institute, Gatchina, Russia
\item \Idef{org97}Physics department, Faculty of science, University of Zagreb, Zagreb, Croatia
\item \Idef{org98}Physics Department, Panjab University, Chandigarh, India
\item \Idef{org99}Physics Department, University of Jammu, Jammu, India
\item \Idef{org100}Physics Department, University of Rajasthan, Jaipur, India
\item \Idef{org101}Physikalisches Institut, Eberhard-Karls-Universit\"{a}t T\"{u}bingen, T\"{u}bingen, Germany
\item \Idef{org102}Physikalisches Institut, Ruprecht-Karls-Universit\"{a}t Heidelberg, Heidelberg, Germany
\item \Idef{org103}Physik Department, Technische Universit\"{a}t M\"{u}nchen, Munich, Germany
\item \Idef{org104}Politecnico di Bari, Bari, Italy
\item \Idef{org105}Research Division and ExtreMe Matter Institute EMMI, GSI Helmholtzzentrum f\"ur Schwerionenforschung GmbH, Darmstadt, Germany
\item \Idef{org106}Rudjer Bo\v{s}kovi\'{c} Institute, Zagreb, Croatia
\item \Idef{org107}Russian Federal Nuclear Center (VNIIEF), Sarov, Russia
\item \Idef{org108}Saha Institute of Nuclear Physics, Homi Bhabha National Institute, Kolkata, India
\item \Idef{org109}School of Physics and Astronomy, University of Birmingham, Birmingham, United Kingdom
\item \Idef{org110}Secci\'{o}n F\'{\i}sica, Departamento de Ciencias, Pontificia Universidad Cat\'{o}lica del Per\'{u}, Lima, Peru
\item \Idef{org111}Shanghai Institute of Applied Physics, Shanghai, China
\item \Idef{org112}St. Petersburg State University, St. Petersburg, Russia
\item \Idef{org113}Stefan Meyer Institut f\"{u}r Subatomare Physik (SMI), Vienna, Austria
\item \Idef{org114}SUBATECH, IMT Atlantique, Universit\'{e} de Nantes, CNRS-IN2P3, Nantes, France
\item \Idef{org115}Suranaree University of Technology, Nakhon Ratchasima, Thailand
\item \Idef{org116}Technical University of Ko\v{s}ice, Ko\v{s}ice, Slovakia
\item \Idef{org117}Technische Universit\"{a}t M\"{u}nchen, Excellence Cluster 'Universe', Munich, Germany
\item \Idef{org118}The Henryk Niewodniczanski Institute of Nuclear Physics, Polish Academy of Sciences, Cracow, Poland
\item \Idef{org119}The University of Texas at Austin, Austin, Texas, United States
\item \Idef{org120}Universidad Aut\'{o}noma de Sinaloa, Culiac\'{a}n, Mexico
\item \Idef{org121}Universidade de S\~{a}o Paulo (USP), S\~{a}o Paulo, Brazil
\item \Idef{org122}Universidade Estadual de Campinas (UNICAMP), Campinas, Brazil
\item \Idef{org123}Universidade Federal do ABC, Santo Andre, Brazil
\item \Idef{org124}University College of Southeast Norway, Tonsberg, Norway
\item \Idef{org125}University of Cape Town, Cape Town, South Africa
\item \Idef{org126}University of Houston, Houston, Texas, United States
\item \Idef{org127}University of Jyv\"{a}skyl\"{a}, Jyv\"{a}skyl\"{a}, Finland
\item \Idef{org128}University of Liverpool, Liverpool, United Kingdom
\item \Idef{org129}University of Science and Techonology of China, Hefei, China
\item \Idef{org130}University of Tennessee, Knoxville, Tennessee, United States
\item \Idef{org131}University of the Witwatersrand, Johannesburg, South Africa
\item \Idef{org132}University of Tokyo, Tokyo, Japan
\item \Idef{org133}University of Tsukuba, Tsukuba, Japan
\item \Idef{org134}Universit\'{e} Clermont Auvergne, CNRS/IN2P3, LPC, Clermont-Ferrand, France
\item \Idef{org135}Universit\'{e} de Lyon, Universit\'{e} Lyon 1, CNRS/IN2P3, IPN-Lyon, Villeurbanne, Lyon, France
\item \Idef{org136}Universit\'{e} de Strasbourg, CNRS, IPHC UMR 7178, F-67000 Strasbourg, France, Strasbourg, France
\item \Idef{org137}Universit\'{e} Paris-Saclay Centre d'Etudes de Saclay (CEA), IRFU, D\'{e}partment de Physique Nucl\'{e}aire (DPhN), Saclay, France
\item \Idef{org138}Universit\`{a} degli Studi di Foggia, Foggia, Italy
\item \Idef{org139}Universit\`{a} degli Studi di Pavia, Pavia, Italy
\item \Idef{org140}Universit\`{a} di Brescia, Brescia, Italy
\item \Idef{org141}Variable Energy Cyclotron Centre, Homi Bhabha National Institute, Kolkata, India
\item \Idef{org142}Warsaw University of Technology, Warsaw, Poland
\item \Idef{org143}Wayne State University, Detroit, Michigan, United States
\item \Idef{org144}Westf\"{a}lische Wilhelms-Universit\"{a}t M\"{u}nster, Institut f\"{u}r Kernphysik, M\"{u}nster, Germany
\item \Idef{org145}Wigner Research Centre for Physics, Hungarian Academy of Sciences, Budapest, Hungary
\item \Idef{org146}Yale University, New Haven, Connecticut, United States
\item \Idef{org147}Yonsei University, Seoul, Republic of Korea
\end{Authlist}
\endgroup
\end{document}